# Narrow Bandgap in β-BaZn$_2$As$_2$ and Its Chemical Origins

Zewen Xiao,† Hidenori Hiramatsu,†‡ Shigenori Ueda,§ Yoshitake Toda,# Fan-Yong Ran,† Jiangang Guo,# Hechang Lei,# Satoru Matsuishi,‡ Hideo Hosono,†‡# and Toshio Kamiya*†‡

† Materials and Structures Laboratory, Tokyo Institute of Technology, Yokohama 226-8503, Japan

‡ Materials Research Center for Element Strategy, Tokyo Institute of Technology, Yokohama 226-8503, Japan

§ Synchrotron X-ray Station at SPring-8, National Institute for Materials Science, Hyogo 679-5148, Japan

# Frontier Research Center, Tokyo Institute of Technology, Yokohama 226-8503, Japan

**ABSTRACT:** β-BaZn$_2$As$_2$ is known to be a *p*-type semiconductor with the layered crystal structure similar to that of LaZnAsO, leading to the expectation that β-BaZn$_2$As$_2$ and LaZnAsO have similar bandgaps; however, the bandgap of β-BaZn$_2$As$_2$ (previously-reported value ~0.2 eV) is one order of magnitude smaller than that of LaZnAsO (1.5 eV). In this paper, the reliable bandgap value of β-BaZn$_2$As$_2$ is determined to be 0.23 eV from the intrinsic region of the temperature dependence of electrical conductivity. The origins of this narrow bandgap are discussed based on the chemical bonding nature probed by 6 keV hard X-ray photoemission spectroscopy, hybrid density functional calculations, and the ligand theory. One origin is the direct As-As hybridization between adjacent [ZnAs] layers, which leads to a secondary splitting of As 4*p* levels and raises the valence band maximum. The other is that the non-bonding Ba 5$d_{x^2-y^2}$ orbitals form unexpectedly deep conduction band minimum (CBM) in β-BaZn$_2$As$_2$ although the CBM of LaZnAsO is formed mainly of Zn 4s. These two origins provide a quantitative explanation for the bandgap difference between β-BaZn$_2$As$_2$ and LaZnAsO.

## INTRODUCTION

Layered mixed-anion compounds including LaCu*Ch*O (*Ch* = S, Se, Te)[1,2], La*T$_M$P$_n$*O (*T$_M$* = Mn, Zn; *Pn* = P, As)[3–5] and La*T$_M$*'*P$_n$*O (*T$_M$*' = Fe, Ni; *Pn* = P, As)[6–8] exhibit a wide variety of electronic phenomena such as wide gap *p*-type semiconduction and superconductivity, making them attractive for a new platform to explore functional materials. These compounds have the general chemical formula *LnMAX* (*Ln* =lanthanide, *M* = transition metal, *A* = chalcogen or pnictogen, *X* = O, F or H), which are called '1111-type' compounds and have the tetragonal ZrCuSiAs-type structure (space group *P*4/*nmm*). In particular, an interesting feature of these compounds is that they have a two-dimensional crystal structure composed of alternating [*MA*] and [*LnX*] layers; the former forms a carrier conduction path and the latter forms a wider bandgap than the conduction layer and behaves like a carrier transport barrier.

On the other hand, similar layered-structure compounds (i.e., composed of a narrow bandgap [*MA*] layer) have also been found; a representative one is called '122-type' compounds expressed by the chemical formula *AeM$_2$Pn$_2$* (*Ae* = alkaline earth). Similar to the 1111-type compounds, the properties of *AeM$_2$Pn$_2$* change drastically if the type of the transition metal *M* is varied. For example, *AeM$_2$Pn$_2$* behave as superconductors (for Fe[9,10] and Ni[11]), ferromagnetic metals (for Co[12]), antiferromagnetic metals (for Cr[13]), diamagnetic metals (for Cu[14]), antiferromagnetic semiconductors (for Mn[15–18]), and non-magnetic semiconductors (for Zn[19]). Most of them crystallize into the tetragonal ThCr$_2$Si$_2$ structure with the space group *I*4/*mmm*.[20] On the other hand, BaZn$_2$As$_2$ has two crystalline phases; the low-temperature orthorhombic phase α-BaZn$_2$As$_2$[21] (α-BaCu$_2$S$_2$-type structure, the space group *Pnma*) and the high-temperature tetragonal phase β-BaZn$_2$As$_2$[22] (the 122-type one, the space group *I*4/*mmm*) (Fig. 1(a)). Recently, (Ba$_{1-x}$K$_x$)(Zn$_{1-y}$Mn$_y$)$_2$As$_2$ was reported to be a good diluted magnetic semiconductor, in which the tetragonal phase is stabilized by doping of 10% K or Mn.[19]

It is known that the anion-anion chemical bonding influences the ground states of 122-type compounds significantly and is intertwined with the formation of the ferromagnetic quantum critical point and superconductivity.[23,24] In our previous work, we roughly determined the bandgap of a β-BaZn2As2 epitaxial film to be ~0.2 eV from optical transmission spectra,[25] which is extremely narrow compared with that of the similar 1111-type compound, LaZnAsO (1.5 eV).[4] Further it has been reported that simple zinc arsenides have much larger bandgaps as well (e.g., 0.99 eV for Zn3As2[26] and 0.98 eV for ZnAs2[27]). Recently, first-principles calculations for the α- and β-BaZn2As2 phases reported that they have complicated and highly-anisotropic electronic structures due to the unusual cation-anion and anion-anion hybridizations;[28] however, it does not provide an explanation for the extreme narrow bandgap of β-BaZn2As2. Further, the previously-reported bandgap value[25] is not so reliable due to the interference of the substrate optical absorption, and no other experimental data on its electronic structure has been provided.

In this work, we determined the bandgap of β-BaZn$_2$As$_2$ from the carrier transport properties and obtained the reliable electronic structure by hard X-ray photoemission spectroscopy (HAXPES). Hybrid density functional theory calculations with Heyd-Scuseria-Ernzerhof (HSE06) functionals provided good agreement with the experimental bandgap and the valence band (VB) HAXPES spectra. We found the anion-anion hybridization between adjacent [ZnAs] layers induces a secondary splitting of the outer orbitals of As atoms and widens the VB. In addition, the $d$ orbitals of the heavy alkaline earth Ba ion form a deep conduction band (CB) minimum (CBM) due to non-bonding nature of the Ba $5d_{x^2-y^2}$ orbitals. These two factors provide a quantitative explanation for the narrow bandgap of β-BaZn$_2$As$_2$.

## EXPERIMENTAL AND COMPUTATIONAL DETAILS

**Synthesis of powder and polycrystalline samples.** Polycrystalline samples of β-BaZn$_2$As$_2$ and LaZnAsO were synthesized through solid-state reactions. First, BaAs and LaAs precursors were synthesized from stoichiometric mixtures of Ba, La, and As pieces / powders, which were sealed in evacuated silica-glass ampules and heated at 700 °C for 20 hours. Then, stoichiometric BaAs, Zn, and As powders (for β-BaZn$_2$As$_2$) and stoichiometric LaAs and ZnO powders (for LaZnAsO) were mixed, respectively, and pre-heated at 700 °C in evacuated silica-glass ampules. The reaction products were grounded and pressed into pellets and annealed at 1000 °C for 20 hours in evacuated silica-glass ampules again. After annealing, the ampule containing LaZnAsO was furnace-cooled to room temperature (RT); while that containing BaZn$_2$As$_2$ was rapidly quenched by dropping into water to stabilize the high-temperature phase, β-BaZn$_2$As$_2$. All the synthesis processes except the sealing, the heating and the cooling processes were carried out in a glove box filled with dry argon gas (dew point < −90 °C, O$_2$ < 1 ppm).

**Structural and electronic measurements.** The phase determination of the obtained samples was carried out by powder x-ray diffraction (PXRD) (D8 ADVANCE, Bruker, using Cu K$\alpha$ rotating anode). PXRD patterns were simulated from literature structures[22,29] using a TOPAS code.[30] The temperature ($T$) dependence of dc electrical conductivity (σ) of the β-BaZn$_2$As$_2$ sample was measured by the conventional four-probe method with a physical property measure measurement system (PPMS, Quantum Design) up to 400 K. HAXPES measurements were performed at the BL15XU undulator beamline ($h\nu$ =5953.4 eV) of SPring-8 at RT,[31] where the binding energy is measured from the Fermi level calibrated using a reference spectrum measured on a gold thin film. Total energy resolution was set to 240 meV, which was confirmed by the Fermi cut-off of the gold thin film.

**Theoretical calculations.** Density functional theory (DFT) and hybrid DFT (HDFT) calculations were performed using the projector augmented plane-wave method implemented in the Vienna *Ab initio* Simulation Program (VASP 5.3).[32] The plane wave cutoff energy was set to 345.9 eV. A 5×5×2 $k$-mesh was used for tetragonal β-BaZn$_2$As$_2$ and LaZnAsO and a 5×5×3 $k$-mesh for trigonal $Ae$Zn$_2$As$_2$ crystals ($Ae$ =Ba, Sr and Ca). For the exchange-correlation functional, we first examined the Perdew-Burke-Ernzerhof (PBE96)[33] generalized gradient approximation (GGA) functionals; however, we found it underestimated the bandgaps of β-BaZn$_2$As$_2$ and LaZnAsO. We then examined hybrid functionals and found the HSE06[34,35] hybrid functionals with the standard mixing parameter of 25% for the exact-exchange term provided reasonable results. Before the electronic structure calculations, variable-cell structure relaxations were performed using the respective functionals.

## RESULTS AND DISCUSSION

**Crystal structure difference between β-BaZn$_2$As$_2$ and LaZnAsO.** Figures 1(a,b) show the crystal structures of β-BaZn$_2$As$_2$ and LaZnAsO, respectively, and Fig. 1(d) shows measured & simulated PXRD patterns for the synthesized BaZn$_2$As$_2$ and LaZnAsO samples. All the diffraction peaks of the BaZn$_2$As$_2$ were reproduced well from the ThCr$_2$Si$_2$-type structure ($I4/mmm$) as reported,[22] substantiating that the metastable β-BaZn$_2$As$_2$ phase was successfully obtained by the rapid quenching method. The crystal structure of LaZnAsO was confirmed to be the ZrCuSiAs-type one ($P4/nmm$) as reported.[29] The literature structural parameters for β-BaZn$_2$As$_2$ and LaZnAsO are summarized in Table S1 (see Supporting Information). The structures of the [ZnAs] layers are almost the same between β-BaZn$_2$As$_2$ and LaZnAsO, in which each (ZnAs$_4$) tetrahedron is connected with four neighboring (ZnAs$_4$) tetrahedra by sharing their edges in the two-dimensional [ZnAs] layer. On the other hand, we can find a distinct difference between these structures in Fig. 1(e), where the coordination numbers of As atoms around an As atom is plotted as a function of As–As distances, $d_{As-As}$. The intralayer As–As distances (i.e., $d_{in}$ and $d_{out}$, as denoted in Figs. 1(a)) are similar between β-BaZn$_2$As$_2$ and LaZnAsO (0.410 – 0.425 nm); whereas, the interlayer As–As distance ($d_{inter}$) for β-BaZn$_2$As$_2$ (0.370 nm) is much shorter than that for LaZnAsO (0.660 nm) due to the different heights between the Ba ion and the LaO layer.

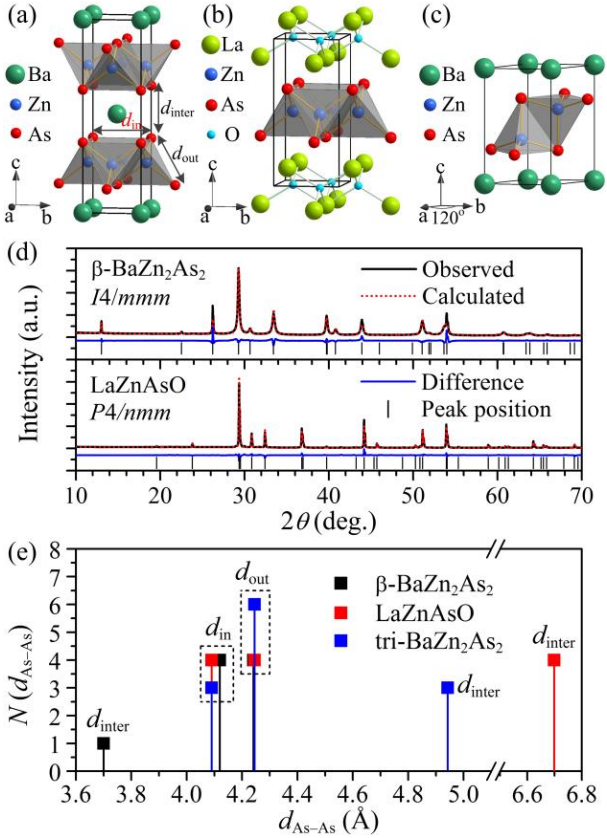

Figure 1. Crystal structures. Schematic illustrations for (a) β-BaZn$_2$As$_2$ ($I4/mmm$), (b) LaZnAsO ($P4/nmm$), and (c) trigonal BaZn$_2$As$_2$ ($P$-$3m1$, denoted as tri-BaZn$_2$As$_2$). The $d_{in}$ (equal to the lattice parameter $a$), $d_{out}$ and $d_{inter}$ parameters are defined as the distances between the nearest As neighbors along the in-plane, the out-of-plane and the interlayer directions, respectively, as denoted in (a). (c) Measured and simulated PXRD patterns for β-BaZn$_2$As$_2$ and LaZnAsO samples. (d) Comparison of coordination structures among β-BaZn$_2$As$_2$, LaZnAsO, and tri-BaZn$_2$As$_2$, which plots the coordination number distribution of As atoms around an As atom as a function of As–As distance ($d_{As-As}$).

**Bandgap determination of β-BaZn$_2$As$_2$.** In our previous work, the optical bandgap of β-BaZn$_2$As$_2$ was evaluated to be ~0.2 eV for an epitaxial film on an MgO substrate.[25] However, the bandgap determination was hindered more or less by the optical absorption of the MgO substrate at 0.05 – 0.20 eV. Here, we further examined the bandgap from the intrinsic region of electrical conductivity ($\sigma$) of a polycrystalline BaZn$_2$As$_2$ sample. The $\sigma$ (in logarithmic coordinate) vs. $1/T$ Arrhenius plot in Fig. 2 exhibits two linear regions between 10 and 50 K and between 190 and 310 K, respectively. The high-$T$ linear region should be attributed to the intrinsic region of $\sigma$ and can be expressed by $\sigma = \sigma_0 \exp(-E_a/k_B T)$, where $\sigma_0$ is the pre-exponential constant, $k_B$ the Boltzmann constant, and $E_a$ the activation energy. The low-$T$ linear region could be attributed e.g. to the variable range hopping (VRH) model, the saturation regime of the ionization of donors, and very shallow donors. For example, we fitted the $\sigma$-$T$ data by a combined expression $\sigma(T) = \sigma_0 \exp(-E_a/k_B T) + \sigma_{vrh}\exp[-(T_0/T)^{1/4}]$, where the latter express the VRH model with constants $\sigma_{vrh}$ and $T_0$,[36] as drawn by red curve. The fit gave intrinsic $E_a$ value of 0.114 eV, which is comparable to those reported for BaMn$_2$P$_2$ (0.07 eV),[15] BaMn$_2$As$_2$ (0.03 eV)[16] and BaMn$_2$Sb$_2$ (0.03 – 0.10 eV)[17,18] (note that these crystals have the same 122-type structure). Thus we conclude that the intrinsic $E_a$ is 0.114 eV and thus the bandgap is $E_g = 2E_a = 0.23$ eV. We would like to note that the change of the slope in the Arrhenius plot around $1/T$ ~ 0.07 seems very abrupt and might suggest a phase transition; however, the fitting result to the combined model shows that the experimental transition is duller than that expected from the theoretical model.

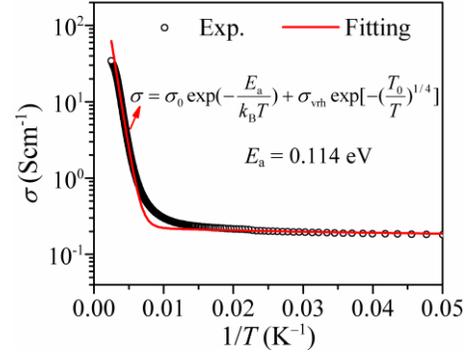

Figure 2. $T$ dependence of $\sigma$ for polycrystalline β-BaZn$_2$As$_2$ sample. The red curve is a fit to $\sigma(T) = \sigma_0\exp(-E_a/k_B T) + \sigma_{vrh}\exp[-(T_0/T)^{1/4}]$.

We also calculated the bandgaps for β-BaZn$_2$As$_2$ and LaZnAsO by DFT calculations. First, we examined the PBE96 GGA functional, but it gave a negative bandgap for β-BaZn$_2$As$_2$ and an underestimated value of 0.56 eV for LaZnAsO, which are consistent with the reported GGA results.[4,25,28] It is caused by a well-known bandgap problem of DFT in which bandgaps are in general underestimated from experimental values. Then, we examined the HSE06 hybrid functional and found that it provided reasonable bandgaps of 0.23 eV for β-BaZn$_2$As$_2$ and 1.38 eV for LaZnAsO, which are fairly close to the experimental values. Besides, HSE06 also gave better structural parameters than PBE96 (see Table S1 in Supporting Information).

**Electronic structure differences between β-BaZn$_2$As$_2$ and LaZnAsO.** For β-BaZn$_2$As$_2$ and LaZnAsO, we had expected that the common [ZnAs] layers should form the bandgap both for the CBM and VBM. However, both the experimental and the calculation results show that the bandgap of β-BaZn$_2$As$_2$ is much smaller than that of LaZnAsO. The similar difference in the bandgap is observed also in the Mn-based compounds (i.e. 0.14 eV for BaMn$_2$P$_2$,[15] 1.7 eV for LaMnPO,[5] 0.06 eV for BaMn$_2$As$_2$,[16] 1.5 eV for LaMnAsO,[5] 0.06 – 0.20 eV for BaMn$_2$Sb$_2$,[17,18] 1.0 eV for LaMnSbO[5]). Here, we discuss the electronic structures of β-BaZn$_2$As$_2$ and LaZnAsO, where we focus on the origins of the narrow bandgap in β-BaZn$_2$As$_2$.

The calculated total and projected densities of states (DOSs) for β-BaZn$_2$As$_2$ and LaZnAsO are shown in Figs. 3(a) and (b). The CB of β-BaZn$_2$As$_2$ between CBM and

CBM + 1.0 eV consists mainly of Ba $5d$ and As $4d$ hybridized with small portion of Zn $4s$ orbitals, which is highly dispersed as seen in the band structure in Fig. 3(c). The CB of LaZnAsO between CBM and CBM + 1.0 eV consists mainly of Zn $4s$ and As $4d$ orbitals with narrower dispersion as seen in Fig. 3(d). Both compounds exhibit long CB tails in the total DOSs near CBM. The origin of the CBM states will be discussed more definitely later on.

The VB of β-BaZn$_2$As$_2$ consists mainly of As $4p$ orbital slightly hybridized with Zn $4p$ and Ba $5d$ orbitals; while that of LaZnAsO consists of As $4p$ orbitals slightly hybridized with Zn $4p$ orbitals in the shallow region (VBM to –3.6 eV), and the O $2p$ orbitals slightly hybridized with As $4p$ and La $5d$ orbitals in the deeper region (–3.6 to –6.0 eV). In particular, β-BaZn$_2$As$_2$ exhibits a long tail structure in the total DOS from –0.6 eV to the VBM.

This characteristic structure is confirmed experimentally by HAXPES shown in Fig. 4(a). Fig. 4(b) magnifies the VB spectra around the VBM, which shows the different VB tail structures between β-BaZn$_2$As$_2$ and LaZnAsO more clearly (note that the energy is measured from VBM and the respective Fermi levels are indicated by '$E_F$' in the figure). LaZnAsO exhibits a linear VB edge just below the VBM along with tail states extending to the bandgap up to 0.2 eV above VBM. On the other hand, β-BaZn$_2$As$_2$ exhibits two linear regions; one appears in the energy region deeper than 0.3 eV, and the other appears in the shallow region to VBM. Although similar tail structures are observed often in defective semiconductors such as amorphous In-Ga-Zn-O[37] and also in the present LaZnAsO data, the calculated DOSs support the experimental data and the above interpretation; i.e., LaZnAsO has a single straight VBM structure while the VBM structure of β-BaZn$_2$As$_2$ is curved as seen in the VB DOS in Fig. 4(c), guaranteeing that the upper linear tail region of Fig. 4(b) (0.35 eV in length) is not defect states but the intrinsic VB states. Besides, the $E_F$ of β-BaZn$_2$As$_2$ and LaZnAsO were 0.04 eV and 0.34 eV above VBM, respectively, which are much lower than the half-bandgap levels, indicating these compounds are doped $p$-type semiconductors.

From the band structures in Figs. 3(c) and (d), we can see that the VB levels are split to four similar energy levels in β-BaZn$_2$As$_2$ at the Z point, which is mainly explained by the intralayer hybridizations between the As atoms. The two bands marked by the red symbols are bonding (σ) and anti-bonding (σ$^*$) σ states of As $4p_z$ orbitals, respectively; while the other two by the blue symbols are bonding (π) and anti-bonding (π$^*$) π states of As ($4p_x$, $4p_y$) orbitals, respectively, as confirmed from the $lm$-decomposed DOSs of As $4p$ orbitals (the fifth panels of Figs. 3(a) and (b)). Each of them is doubly degenerated at the Z point, while splits to two bands along the $k$ vector moving to the Γ point because the adjacent [ZnAs] layers interact with each other and form a bonding and an anti-bonding states as indicated by σ$^*_1$ and σ$^*_2$ for example in Fig. 3(c). For LaZnAsO, the degenerated levels of As $4p$ at the Z point are the same as β-BaZn$_2$As$_2$; while, each band does not exhibit a significant energy split along the Z – Γ direction (Fig. 3(d)), which is strikingly different from β-BaZn$_2$As$_2$.

Comparing the different VB structures, the most significant difference near the VBM is found in the σ$^*_2$ band of β-BaZn$_2$As$_2$ formed due to the direct interlayer As $4p_z$ – As $4p_z$ anti-bond (see Fig. 5(a)), which is highly dispersed and the maximum energy level exceeds the π$^*_2$ band by 0.36 eV at the Γ point; i.e., As $4p_z$ mainly forms the $lm$-decomposed DOS of the shallow VB region as seen in the fifth panel of Fig. 3(b). This explains the tail structures observed in the total DOS and the HAXPES VB spectra (Fig. 4) for β-BaZn$_2$As$_2$.

The different VBM dispersions are understood more clearly by the relative effective hole masses ($m_h^*/m_0$). The obtained $m_h^*/m_0$ values are 0.094 along Γ – Z and 0.88 along Γ – X for β-BaZn$_2$As$_2$, which are much smaller than those of LaZnAsO (0.72 and 4.32, respectively). This result suggests that β-BaZn$_2$As$_2$ should exhibit better hole transport along the $c$-axis than that in the $a$-$b$ plane and also that β-BaZn$_2$As$_2$ should exhibit much better hole transport than LaZnAsO.

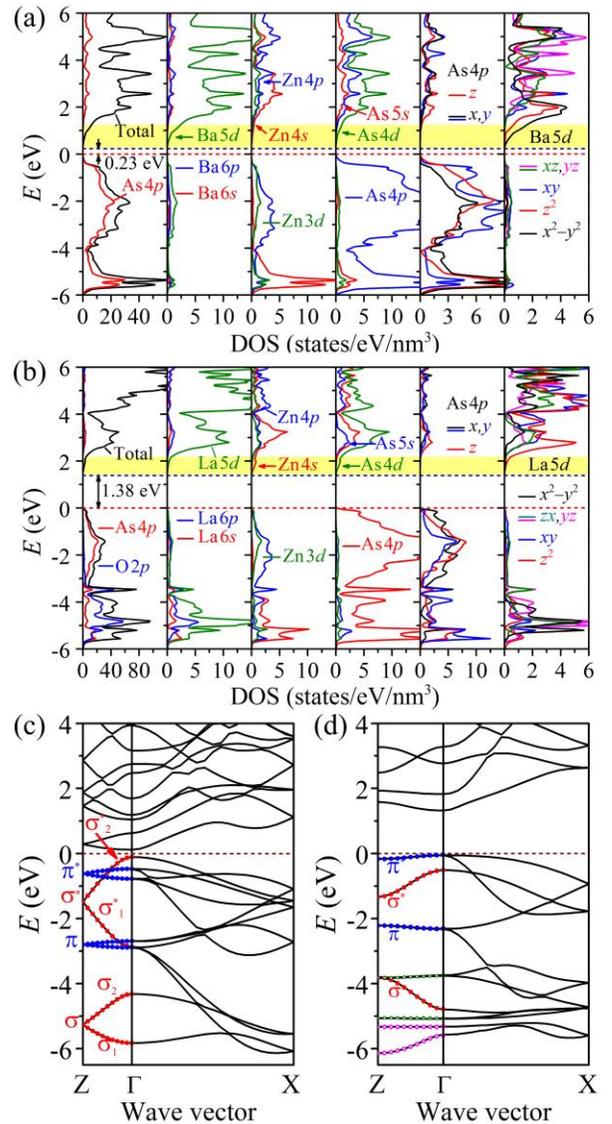

Figure 3. (a and b) Total, projected, and *lm*-decomposed densities of states (DOS) of (a) β-BaZn$_2$As$_2$ and (b) LaZnAsO. The red and blue lines mark the valence band maximum (VBM) and the conduction band minimum (CBM), respectively. (c and d) Band structures of (c) β-BaZn$_2$As$_2$ and (d) LaZnAsO. The dashed lines mark the VBMs.

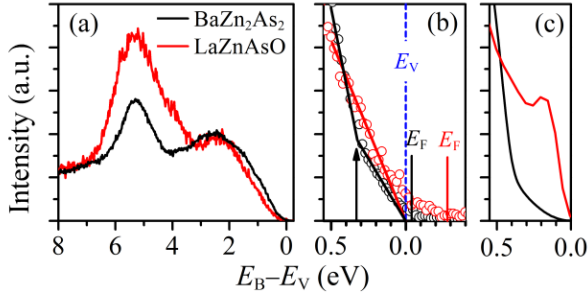

Figure 4. Valence band (VB) spectra of β-BaZn$_2$As$_2$ and LaZnAsO. (a) HAXPES VB spectra. (b) Magnified HAXPES VB spectra near VBM. (c) Calculated HSE06 DOSs. The binding energy ($E_B$) is aligned with respect to the VBM level ($E_V$).

**Origins of narrow bandgap of β-BaZn$_2$As$_2$ I: interlayer As–As hybridization.** As discussed above, the Γ point splitting of the VB in β-BaZn$_2$As$_2$ (denoted as secondary splitting hereafter) is caused by the direct hybridization between interlayer As atoms (schematically shown in Fig. 5(a)) due to the relatively short $d_{inter}$ (0.370 nm). To further confirm this model, we performed HSE06 calculations for several hypothetical structures of β-BaZn$_2$As$_2$ with fixed [ZnAs] layers and varied $d_{inter}$ values, as illustrated in Fig. 5(a). The obtained bandgap vs. $d_{inter}$ relation is summarized in Fig. 5(b), and those band structures and DOSs are shown in Figs. 5(c) and (d), respectively. First, we slightly decreased the $d_{inter}$ from 0.370 to 0.360 nm. The secondary splitting of As 4$p$ became wider because of the larger hybridization of the interlayer As atoms, which raised the VBM energy level as shown in the left panel in Fig. 5(c). As a result, the bandgap value is decreased to 0.04 eV. When the $d_{inter}$ was increased to 0.380 nm, the secondary splitting of As 4$p$ became smaller due to the reduced interlayer hybridizations, which led to a lowered VBM and an increased bandgap of 0.40 eV. By further increasing the $d_{inter}$ to 0.400 nm, the secondary splitting became further smaller and the $\sigma^*_2$ band did not pass across the $\pi^*_2$ band anymore (the right panel of Fig. 5(c)), which made the VBM structure more similar to that of LaZnAsO. Correspondingly, the "tail" structure near VBM faded away completely as seen in Fig. 5(d). The resulted bandgap was further increased to 0.66 eV. We confirmed that further increasing $d_{inter}$ did not increase the bandgap significantly. From the $d_{inter}$ dependence of bandgap summarized in Fig. 5(b), we can conclude that the interlayer hybridization between the As 4$p_z$ orbitals of adjacent layers is an origin of the narrow bandgap in β-BaZn$_2$As$_2$, which decreases the bandgap by ~0.43 eV (the bandgap difference between the actual structure ($d_{inter}$ = 0.370 nm) and that with $d_{inter}$ being 0.400 nm). Here, we should note that the $d_{inter}$ value in β-BaZn$_2$As$_2$ is shorter than that in LaZnAsO, but much longer than As–As bonds in single-bonded As$_2$-dimers (0.240 – 0.260 nm)[38] in such as CaNi$_2$As$_2$.[39] This means that the interlayer hybridization of the As 4$p_z$ orbitals in β-BaZn$_2$As$_2$ is much weaker than the As$_2$-dimers and is of an intermediate case.

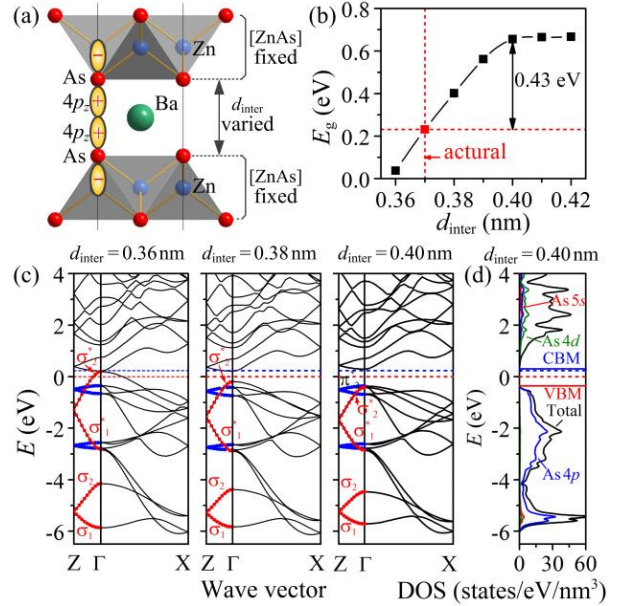

Figure 5. (a) Schematic illustration of hypothetical structures of β-BaZn$_2$As$_2$ with fixed [ZnAs] layer structures and varied $d_{inter}$ values. (b) Calculated bandgap of β-BaZn$_2$As$_2$ as a function of $d_{inter}$. (c) Band structures of hypothetical β-BaZn$_2$As$_2$ structures with $d_{inter}$ being 0.36, 0.38 and 0.40 nm. (d) Total and projected DOSs of hypothetical β-BaZn$_2$As$_2$ with $d_{inter}$ = 0.40 nm. In (c) and (d), the energies are aligned to the energy levels of the Zn 3$d$ orbitals, and the dashed red and blue lines mark the VBM and the CBM of the experimental β-BaZn$_2$As$_2$ structure ($d_{inter}$ = 0.37 nm) for comparison.

**Origins of narrow bandgap of β-BaZn$_2$As$_2$ II: non-bonding Ba 5$d_{x^2-y^2}$ orbital.** As discussed above for the PDOS in Fig. 3(a), the Ba 5$d$ orbitals contribute largely to the CB, in particular to the CBM structure. To examine the role of the Ba 5$d$ orbitals on the narrow bandgap, we further calculated *lm*-decomposed DOSs for the five Ba 5$d$ orbital (i.e. $d_{xy}$, $d_{xz}$, $d_{yz}$, $d_{x^2-y^2}$ and $d_{z^2}$) as shown in the rightmost panel of Fig. 3(a). It is known well that the energy levels of outer $d$ orbitals are split by the coordinating ligands, whose energy splits are understood qualitatively from the ligand symmetry ($D_{4h}$ around Ba, as shown in Fig. 6(a)) by the group theory. The energy level of the edges of Ba 5$d$ orbitals derived from the ligand theory and the *lm*-decomposed DOSs is schematically shown in Fig. 6(c). As known from the *lm*-decomposed DOSs, the $d_{x^2-y^2}$ has the lowest energy among the five $d$ orbitals, and is even lower than that of Zn 4$s$ orbital (the dashed line in Fig. 6(c)), and forms the CBM. As drawn in Fig. 6(a), the $d_{x^2-y^2}$ wave function extends to the interstitial spaces between the neighboring As 4$p$ orbitals, minimizes the charge overlap with the As electrons, lowers its energy level due to the small Coulomb repulsion, and forms the deep CBM. Further, as the Ba 5$d_{x^2-y^2}$ – As 4$p$ states are non-bonding states due to the point-group symmetry and the restriction of the translational symmetry at the Ba site, which diminishes the energy upshift due to the anti-bonding interaction between the Ba $5d_{x^2-y^2}$ and As $4p$ orbitals and also contributes to the formation of the deep CBM. This is similar to the case of superdegeneration observed *e.g.* in cubic perovskites; for example, the bandgaps of $GeO_2$ are >6 eV, but that in cubic $SrGeO_3$ is reduced to 2.7 eV due to the non-bonding nature of Ge 4s.[40]

In contrast, LaZnAsO has a lower ligand symmetry of $C_{4v}$ around La (see Fig. 6(b)), and each La $5d$ orbital cannot avoid hybridization with O $2p$ and As $4p$ orbitals, where anti-bonding interaction raises the CBM. The energy spilt of La $5d$ orbitals is much narrower, as seen from the *lm*-decomposed DOSs in the rightmost panel of Fig. 3(b) and the derived schematic energy diagram in Fig. 6(d). Besides, the energy levels of La $5d$ orbitals are also up-shifted significantly and are higher than the Zn $4s$ level due to the anti-bonding interaction between the La $5d$ and the O $2p$ and As $4p$ orbitals. As a result, La $5d$ orbitals do not contribute to the CBM structure, and the bandgap is mainly formed in the [ZnAs] layer (note that the bandgap of a single [ZnAs] layer is 1.48 eV. See Fig. S1 in Supporting Information). From the above discussion, we conclude that the non-bonding Ba $5d_{x^2-y^2}$ orbital forms the deep CBM and contributes to the narrow bandgap of β-$BaZn_2As_2$.

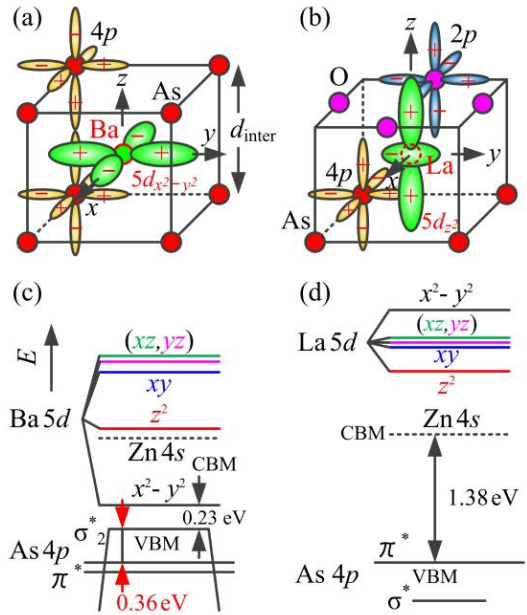

Figure 6. (a,b) Ligand geometries of (a) $D_{4h}$ around Ba in β-$BaZn_2As_2$ and (b) $C_{4v}$ around La in LaZnAsO. The wave functions of the As $4p$ orbitals, the O $2p$ orbitals and the lowest $5d$ orbitals of Ba and La are also shown schematically. (c,d) Schematic energy diagrams near the VBM and the CBM for (c) β-$BaZn_2As_2$ and (d) LaZnAsO. The energy levels are derived from the *lm*-decomposed DOSs in Figs. 3(a,b) and aligned by Zn $3d$ levels.

**Large bandgaps in trigonal 122-type pnictide semiconductors.** From the above discussion, we concluded that the two origins (the direct hybridization between the interlayer As atoms and the nonbonding state of Ba $5d_{x^2-y^2}$) cause the narrow bandgap in 122-type pnictides due to the local $D_{4h}$ symmetry. The latter origin suggests that the energy level of Ba $5d$ is affected largely by the ligand field and the local symmetry; i.e., lower local symmetry would raise the Ba $5d$ levels and widen the bandgap of 122-type compounds.

Here, we examined the electronic structure of a hypothetical trigonal $BaZn_2As_2$ (Fig. 1(c), denoted as tri-$BaZn_2As_2$ hereafter) because a trigonal 122-type structure with $D_{3d}$ local symmetry (space group $P\bar{3}m1$) has been reported for $AeT_{M2}Pn_2$.[15,41,42] Due to the trigonal symmetry in tri-$BaZn_2As_2$, each $[ZnAs_4]$ tetrahedron connects with 3 neighboring $[ZnAs_4]$ tetrahedra by sharing their edges, which is different from the tetragonal β-$BaZn_2As_2$ and LaZnAsO (4 neighboring $[ZnAs_4]$ tetrahedra). Compared with β-$BaZn_2As_2$, the tri-$BaZn_2As_2$ has similar intralayer As–As distances (i.e., $d_{in}$ and $d_{out}$) but a much larger interlayer As–As distance ($d_{inter}$ = 0.494 nm) as shown in Fig. 1(e).

The calculated DOSs and band structure are shown in Figs. 7(a) and (b), respectively. Just as expected, the bandgap of tri-$BaZn_2As_2$ is increased significantly to 0.92 eV. This bandgap increase is understandable based on the above discussion. The As $4p$ orbitals do not exhibit the secondary splitting and do not raise the VBM unlikely observed in β-$BaZn_2As_2$ (see Fig. 7(b)) due to the large $d_{inter}$ value. On the other hand, due to the $D_{3d}$ symmetry around the Ba atoms, all the five Ba $5d$ orbitals hybridize with As $4p$ orbitals (seen in Fig. 7(c)) and have almost the similar energy levels, as can be seen in the *lm*-decomposed DOSs in the rightmost panel of Fig. 7(a) and as summarized in the energy diagram in Fig. 7(d). Consequently, the energy levels of Ba $5d$ orbitals are pushed up to a higher energy than Zn $4s$ so that the CBM mainly consist of Zn $4s$ orbitals hybridized with Ba $5d$ and As $4d$ orbitals.

We further examined tri-$SrZn_2As_2$ and tri-$CaZn_2As_2$ and found that they have similar electronic structures as that of tri-$BaZn_2As_2$ (see Fig. S2 in Supporting Information). Their total energies (Fig. 7(e)) are consistent with experimental results; i.e., the tri-$BaZn_2As_2$ structure has a higher energy than the tetragonal β-$BaZn_2As_2$, while the trigonal structures are more stable for $SrZn_2As_2$ and $CaZn_2As_2$. The calculated bandgaps are 1.12 and 1.27 eV, respectively. The order of the bandgap increase from tri-$BaZn_2As_2$ to tri-$SrZn_2As_2$, and finally to tri-$CaZn_2As_2$ can be understood from the increase in the energy levels of the outer $d$ orbitals from Ba to Ca. This result indicates that the difference in the $d$ energy levels is 0.35 eV among Ba, Sr, and Ca in the trigonal structures.

We also calculated the $m_h^*/m_o$ from the VBM band dispersions. The $m_h^*/m_o$ values are 0.522 along the Γ – Z and 0.486 along the Γ – X for tri-$SrZn_2As_2$, and 0.508 along the Γ – Z and 0.551 along the Γ – X for tri-$CaZn_2As_2$. These small $m_h^*/m_o$ values suggest that tri-$BaZn_2As_2$ to tri-$CaZn_2As_2$ could be good *p*-type semiconductors with high hole mobilities.

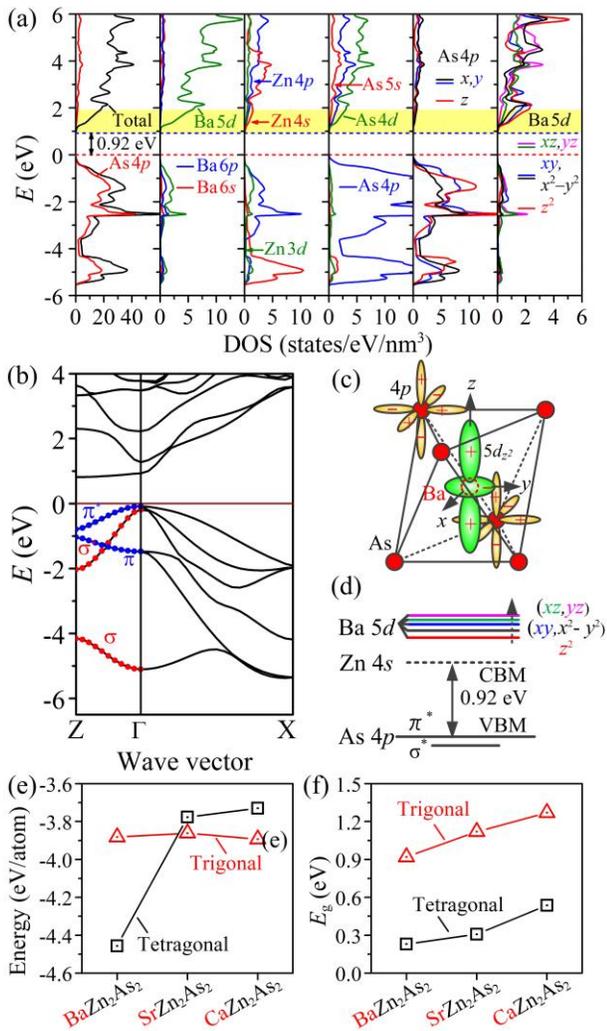

Figure 7. (a) Total, projected and *lm*-decomposed DOSs, and (b) band structure of tri-BaZn$_2$As$_2$. (c) Ligand geometry of $D_{4h}$ around Ba in tri-BaZn$_2$As$_2$. The wave functions of the As 4$p$ orbitals and the Ba 5$d_{z^2}$ orbital are also shown schematically. (d) Schematic energy diagram near the VBM and the CBM of tri-BaZn$_2$As$_2$. (e) Total energy per atom and (f) bandgaps of $Ae$Zn$_2$As$_2$ ($Ae$ = Ba, Sr, and Ca) with tetragonal and trigonal structures with the HSE06 hybrid functional.

## CONCLUSIONS

The bandgap of β-BaZn$_2$As$_2$ was determined to be 0.23 eV from the intrinsic region of the temperature dependence of electrical conductivity. This bandgap value is much smaller than that of a family compound LaZnAsO with the similar crystal structure. The HSE06 calculations, which reproduce the experimental bandgap values and the VB structures of β-BaZn$_2$As$_2$ and LaZnAsO, revealed that the extremely narrow bandgap in β-BaZn$_2$As$_2$ originates from two reasons; one is the pushed-up VBM which is primarily composed of anti-bonding states of As 4$p_z$ orbitals in the adjacent layers, and the other is the extremely low-lying energy level of the non-bonding Ba 5$d_{x^2-y^2}$ orbitals, which is lower even than Zn 4$s$ states. It is believed that the present results provide a clue to better understanding of the evolution of band structure and flexible control of bandgap and carrier transport in versatile pnictide compounds.

## ASSOCIATED CONTENT

**Supporting Information**

Literature and calculated lattice parameters, electronic structures of a single [ZnAs] layer, tri-CaZn$_2$As$_2$ and tri-CaZn$_2$As$_2$. This material is available free of charge via the Internet at http://pubs.acs.org.

## AUTHOR INFORMATION

**Corresponding Author**

* kamiya.t.aa@m.titech.ac.jp

**Notes**

The authors declare no competing financial interests.

## ACKNOWLEDGMENT

This work was conducted under Tokodai Institute for Element Strategy (TIES) funded by MEXT Elements Strategy Initiative to Form Core Research Center. The HAXPES experiments were performed with the approval of NIMS Synchrotron X-ray Station (Proposal Nos. 2012B4612, 2013A4714, 2013A4715, 2013B4703, and 2013B4704). One of the authors (SU) would like to thank HiSOR, Hiroshima University and JAEA at SPring-8 for development of HAXPES at BL15XU of SPring-8.

## REFERENCES

(1) Ueda, K.; Inoue, S.; Hirose, S.; Kawazoe, H.; Hosono, H. *Appl. Phys. Lett.* **2000**, 77, 2701–2703.
(2) Hiramatsu, H.; Ueda, K.; Ohta, H.; Hirano, M.; Kamiya, T.; Hosono, H. *Appl. Phys. Lett.* **2003**, 82, 1048–1050.
(3) Kayamura, K.; Hiramatsu, H.; Hirano, M.; Kawamura, R.; Yanagi, H.; Kamiya, T.; Hosono, H. *Phys. Rev. B* **2007**, 76, 195325.
(4) Kayamura, K.; Kawamura, R.; Hiramatsu, H.; Yanagi, H.; Hirano, M.; Kamiya, T.; Hosono, H. *Thin Solid Films* **2008**, 516, 5800–5804.
(5) Kayanuma, K.; Hiramatsu, H.; Kamiya, T.; Hirano, M.; Hosono, H. *J. Appl. Phys.* **2009**, 105, 073903.
(6) Kamihara, Y.; Hiramatsu, H.; Hirano, M.; Kawamura, R.; Yanagi, H.; Kamiya, T.; Hosono, H. *J. Am. Chem. Soc.* **2006**, 128, 10012–10013.
(7) Kamihara, Y.; Watanabe, T.; Hirano, M.; Hosono, H. *J. Am. Chem. Soc.* **2008**, 130, 3296–3297.
(8) Watanabe, T.; Yanagi, H.; Kamihara, Y.; Kamiya, T.; Hirano, M.; Hosono, H. *J. Solid State Chem.* **2008**, 181, 2117–2120.
(9) Rotter, M.; Tegel, M.; Johrendt, D. *Phys. Rev. Lett.* **2008**, 101, 107006.
(10) Katase, T.; Hiramatsu, H.; Yanagi, H.; Kamiya, T.; Hirano, M.; Hosono, H. *Solid State Commun.* **2009**, 149, 2121–2124.
(11) Bauer, E. D.; Ronning, F.; Scott, B. L.; Thompson, J. D. *Phys. Rev. B* **2008**, 78, 172504.
(12) Singh, D. J.; Sefat, A. S.; McGuire, M. A.; Sales, B. C.; Mandrus, D. *Phys. Rev. B* **2009**, 79, 094429.
(13) Sefat, A. S.; Singh, D. J.; Jin, R.; McGuire, M. A.; Sales, B. C.; Mandrus, D. *Phys. Rev. B* **2009**, 79, 024512.
(14) Anand, V. K.; Perera, P. K.; Pandey, A.; Goetsch, R. J.; Kreyssig, A.; Johnston, D. C. *Phys. Rev. B* **2012**, 85, 214523.
(15) Brock, S. L.; Greedan, J. E.; Kauzlarich, S. M. *J. Solid State Chem.* **1994**, 113, 303–311.
(16) Singh, Y.; Ellern, A.; Johnston, D. C. *Phys. Rev. B* **2009**, 79, 094519.

(17) Wang, H.F.; Cai, K. F.; Li, H.; Wang, L.; Zhou, C. W. *J. Alloys Compd.* **2009**, 477, 519–522.

(18) An, J.; Sefat, A. S.; Singh, D. J.; Du, M. -H. *Phys. Rev. B* **2009**, 79, 075120.

(19) Zhao, K.; Deng, Z.; Wang, X. C.; Han, W.; Zhu, J. L.; Li, X.; Liu, Q. Q.; Yu, R. C.; Goko, T.; Frandsen, B.; Liu, L.; Ning, F.; Uemura, Y. J.; Dabkowska, H.; Luke, G. M.; Luetkens, H.; Morenzoni, E.; Dunsiger, S. R.; Senyshyn, A.; Böni, P.; Jin, C. Q. *Nat. Commun.* **2013**, 4, 1442.

(20) Just, G.; Paufler, P. *J. Alloys Compd.* **1996**, 232, 1–25.

(21) Klüfers, P.; Mewis, A. *Z. Naturforsch.* **1978**, 33b, 151–155.

(22) Hellmann, A.; Löhken, A.; Wurth, A.; Mewis, A. *Z. Naturforsch.* **2007**, 62b, 155–161.

(23) Kawasaki, S.; Tabuchi, T.; Wang, X. F.; Chen, X. H.; Zheng, G. -Q. *Supercond. Sci. Technol.* **2010**, 23, 054004.

(24) Jia, S.; Jiramongkolchai, P.; Suchomel, M. R.; Toby, B. H.; Checkelsky, J. G.; Ong, N. P.; Cava R. *J. Nat. Phys.* **2011**, 7, 207–210.

(25) Xiao, Z.; Ran, F. -Y.; Hiramatsu, H.; Matsuishi, S.; Hosono, H.; Kamiya, T. *Thin Solid Films* **2013**, 559, 100–104.

(26) Misiewicz, J.; Pawlikowski, J. M. *Solid State Commun.* **1979**, 32, 687–690.

(27) Mudryi, A. V.; Patuk, A. I.; Shakin, I. A.; Kalmykov, A. E.; Marenkin, S. F.; Raukhman, A. M. *Mater. Chem. Phys.* **1996**, 44, 151–155.

(28) Shein, I. R.; Ivanovskii, A. L. *J. Alloys Compd.* **2014**, 583, 100–105.

(29) Lincke, H.; Glaum, R.: Dittrich, V.; Möller, M. H.; Pöttgen, R. *Z. Anor. Allg. Chem.* **2009**, 635, 936–941.

(30) *TOPAS*, Version 4.2, Bruker AXS: Karlsruhe, Germany, 2009.

(31) Ueda, S.; Katsuya, Y.; Tanaka, M.; Yoshikawa, H.; Yamashita, Y.; Ishimaru, S.; Matsushita, Y.; Kobayashi, K. *AIP Conf. Proc.* **2010**, 1234, 403–406.

(32) Kresse, G.; Furthmüller, J. *Phys. Rev. B* **1996**, 54, 11169-11186.

(33) Perdew, J. P.; Burke, K.; Ernzerhof, M. *Phys. Rev. Lett.* **1996**, 77, 3865–3868.

(34) Heyd, J.; Scuseria, G. E.; Ernzerhof, M. *J. Chem. Phys.* **2003**, 118, 8207–8215.

(35) Heyd, J.; Scuseria, G. E.; Ernzerhof, M. *J. Chem. Phys.* **2006**, 124, 219906.

(36) Mott, N.F. *Phil. Mag.* **1969**, 19, 835–852.

(37) Nomura, K.; Kamiya, T.; Yanagi, H.; Ikenaga, E.; Yang, K.; Kobayashi, K.; Hirano, M.; Hosono, H. *App. Phys. Lett.* **2008**, 92, 202117.

(38) Hoffmann, R.; Zheng, C. *J. Phys. Chem.* **1985**, 89, 4175–4181.

(39) Poble, R.; Frankovsky, R.; Johrendt, D. *Z. Naturforsch.* **2013**, 68b, 581–586.

(40) Mizoguchi, H.; Kamiya, T.; Matsuishi, S.; Hosono, H. *Nat. Commun.* **2011**, 2, 470.

(41) Klüfers, P.; Mewis, A. *Z. Naturforsch.* **1977**, 32b, 753–756.

(42) Mewis, A. *Z. Naturforsch.* **1980**, 35b, 939–941.

**SYNOPSIS TOC**

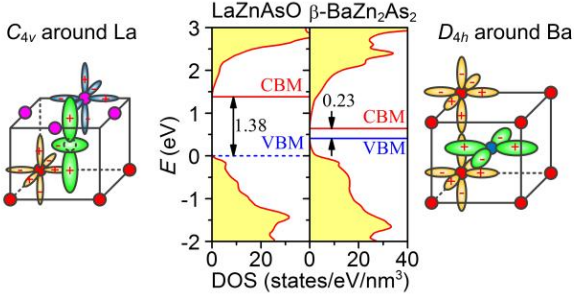